\documentclass[paper=a4,fontsize=12pt,twocolumn]{article}

\usepackage[T1]{fontenc}
\usepackage{fourier}
\usepackage[english]{babel}
\usepackage{fontenc}
\usepackage[left=25mm, top=25mm, right=25mm, bottom=30mm]{geometry}
\usepackage{graphicx}
\usepackage[utf8]{inputenc}
\usepackage{amsfonts}
\usepackage{amsmath}
\usepackage{amssymb}
\usepackage{amsthm}
\usepackage{mathbbol}
\usepackage{mathtools}
\usepackage{xcolor}
\usepackage{epstopdf}
\usepackage{float}
\usepackage{indentfirst}
\usepackage{multicol}
\usepackage{booktabs}
\usepackage{ragged2e}
\usepackage[hidelinks]{hyperref}
\usepackage{tikz}

\title{Black Hole Thermodynamics: Established Results, Unresolved Paradoxes, and Speculative Resolutions}
\author{
    Ricardo Bulcão Valente Ferrari\footnote{Centro de Rádio Astronomia e Astrofísica Mackenzie, São Paulo, Brazil\\ \texttt{ricardobvferrari@gmail.com}}
    \and
    S. B. Soltau\footnote{Department of Physics, Universidade Federal de Alfenas, Minas Gerais, Brazil\\ \url{samuel.soltau@unifal-mg.edu.br}}
    }
\date{June 02, 2025}

\begin{document}

\twocolumn[
\begin{@twocolumnfalse}
\maketitle
\begin{abstract} \noindent
Between 1972 and 1975, Jacob Bekenstein proposed that black holes possess entropy proportional to their horizon area, and Stephen Hawking derived this relationship from semiclassical quantum field theory in curved spacetime, predicting thermal radiation from black holes. These developments established black hole thermodynamics as a formal framework connecting general relativity, quantum mechanics, and statistical physics. However, this synthesis rests on approximations whose validity remains unproven in regimes where quantum gravitational effects become important. This article provides a detailed overview of the historical development from 1972 to 1975 and surveys modern proposals, such as the holographic principle and gravitational path integrals. We highlight persistent theoretical challenges, including the information paradox, the trans-Planckian problem, backreaction effects, and the absence of experimental verification. The work concludes by identifying which aspects of black hole thermodynamics are well-established and which remain speculative or fundamentally incomplete.
\end{abstract}
\end{@twocolumnfalse}
]

\section{Introduction}

Since the first exact solutions to Einstein's field equations~\cite{Einstein1916}, such as the Schwarzschild~\cite{Schwarzschild1916} metric describing non-rotating black holes and the Kerr~\cite{Kerr1963} metric describing rotating ones, black holes have presented profound conceptual challenges at the intersection of thermodynamics, quantum mechanics, and general relativity. In classical general relativity, a black hole is characterized entirely by its mass, angular momentum, and electric charge --- the famous ``no-hair'' theorem. Any additional information about matter that formed the black hole or subsequently fell into it appears to be irretrievably lost, seemingly violating the second law of thermodynamics which requires that total entropy never decreases in isolated systems.

The early 1970s witnessed a remarkable series of theoretical developments attempting to resolve this tension. Jacob Bekenstein, motivated by thought experiments involving entropy-carrying systems falling into black holes, proposed in 1972 that black holes themselves possess entropy proportional to the area of their event horizon~\cite{Bekenstein1972}. This proposal was initially controversial and faced skepticism from leading physicists including Stephen Hawking, who argued that without an emission mechanism, black holes could not possess genuine thermodynamic properties~\cite{Bardeen1973,Hawking1973}. However, Hawking's subsequent application of quantum field theory to curved spacetime backgrounds led to his 1974 prediction that black holes emit thermal radiation~\cite{Hawking1974}, providing a physical mechanism supporting Bekenstein's conjecture.

This apparent synthesis, however, should not be mistaken for a complete resolution of the underlying issues. The derivation of Hawking radiation relies on the \emph{semiclassical approximation}, treating spacetime as a classical curved background on which quantum fields evolve. This approximation is uncontrolled in the sense that we cannot rigorously estimate the magnitude of neglected quantum gravitational corrections, and it certainly breaks down during the final stages of black hole evaporation when the black hole mass approaches the Planck scale. Furthermore, Hawking's original calculation suggested that information falling into a black hole is destroyed~\cite{Hawking1976}, contradicting the unitary evolution required by quantum mechanics. This \emph{information paradox} has remained a central challenge for nearly five decades.

This article provides a technically detailed examination of the historical development of black hole thermodynamics from 1972 to 1975, presenting the key results with dimensionally consistent equations and explicit identification of the physical assumptions underlying each derivation. We then survey modern developments, distinguishing carefully between results that command broad consensus within the theoretical physics community and those that remain speculative or contested. Throughout, we maintain a critical perspective that acknowledges both the remarkable achievements of this theoretical framework and its substantial limitations.

Our goal is not merely to present the canonical results, but to equip the reader with an understanding of what we genuinely know, what we think we know but cannot yet prove rigorously, and what remains fundamentally mysterious. Only by maintaining this epistemological clarity can we appreciate both the genuine progress that has been made and the profound challenges that remain for future research in quantum gravity.

\section{Theoretical Background and Conceptual Foundations}

Before examining Bekenstein's and Hawking's specific contributions, we must establish the theoretical landscape in which they worked and clarify the conceptual framework underlying black hole thermodynamics. This section addresses three foundational questions: What is the classical thermodynamic analogy? What physical regime does semiclassical gravity describe? What are the fundamental limitations of this framework?

\subsection{The Laws of Black Hole Mechanics}

In the early 1970s, Bardeen, Carter, and Hawking~\cite{Bardeen1973} established four laws governing black hole dynamics in classical general relativity that bear striking formal resemblance to the laws of thermodynamics. For a stationary black hole characterized by mass $M$, angular momentum $J$, and electric charge $Q$, the first law relates infinitesimal changes in these quantities:
\begin{equation}\label{eq:first_law_bh}
\delta M = \frac{\kappa}{8\pi G} \delta A + \Omega \, \delta J + \Phi \, \delta Q,
\end{equation}
where $\kappa$ is the surface gravity (constant over the event horizon for stationary black holes), $A$ is the horizon area, $\Omega$ is the angular velocity, and $\Phi$ is the electric potential at the horizon. This equation is structurally identical to the first law of thermodynamics, $\delta E = T\delta S + \text{work terms}$, suggesting the correspondences $\kappa \leftrightarrow T$ and $A \leftrightarrow S$.

However, this analogy was initially regarded as purely formal for a compelling reason: classical black holes do not emit anything. A truly thermodynamic system in thermal equilibrium must exchange energy with its environment, reaching the same temperature. A classical black hole at temperature $T$ would need to radiate, but classical general relativity permits no such radiation mechanism. Hawking himself initially argued strongly against interpreting the analogy as physical~\cite{Hawking1973}, stating that it would be ``tragical'' if black holes actually had temperature, as this would lead to conceptual contradictions within classical theory.

The \emph{zeroth law} states that surface gravity is constant over the event horizon of a stationary black hole, analogous to uniform temperature in thermal equilibrium. The \emph{second law}, proven by Hawking for classical processes, states that horizon area never decreases, $\delta A \geq 0$, analogous to entropy increase. The \emph{third law} states that surface gravity cannot be reduced to zero in a finite number of steps, analogous to the unattainability of absolute zero temperature.

\subsection{The Semiclassical Approximation}

The theoretical framework underlying Hawking's derivation of black hole radiation is \emph{semiclassical gravity}, which treats spacetime as a classical curved manifold described by Einstein's equations while matter fields are quantized. Specifically, Einstein's equations are modified to:
\begin{equation}\label{eq:semiclassical_einstein}
G_{\mu\nu} = \frac{8\pi G}{c^4} \langle T_{\mu\nu} \rangle,
\end{equation}
where the right-hand side now contains the expectation value of the quantum stress-energy tensor operator rather than a classical stress-energy tensor. This approximation is valid when quantum fluctuations in the metric are negligible compared to the background curvature. A quantitative criterion is that components of the Riemann curvature tensor satisfy \(|R_{\mu\nu\rho\sigma}| L_P^2 \ll 1\), where \(L_P = \sqrt{\hbar G/c^3} \approx 1.6 \times 10^{-35}\) m is the Planck length. For a Schwarzschild black hole of mass \(M\), the typical curvature near the horizon scales as \(1/R_s^2\), with \(R_s = 2GM/c^2\) the Schwarzschild radius. Hence, the condition reduces to \((L_P/R_s)^2 \sim (m_P/M)^2 \ll 1\), where \(m_P = \sqrt{\hbar c/G} \approx 2.2 \times 10^{-8}\) kg is the Planck mass. Consequently, the semiclassical treatment is reliable only for black holes with \(M \gg m_P\).

For macroscopic black holes with masses much larger than the Planck mass, this approximation appears well-justified near and outside the event horizon. However, several critical limitations must be acknowledged:

\textbf{First}, the validity condition \( (m_P/M)^2 \ll 1 \) is progressively violated as the black hole evaporates and loses mass. Quantum gravitational corrections, organized as a series in powers of \( L_P^2 R \sim (m_P/M)^2 \), become significant well before the Planck mass is reached. Estimates suggest the semiclassical description becomes unreliable for masses \( M \lesssim 10\, m_P \), a regime where the final stages of evaporation are governed by unknown quantum gravity dynamics. The oft-cited breakdown at \( M \sim m_P \) is a conservative oversimplification; the transition to a fully quantum-gravitational regime is gradual and begins earlier.

\textbf{Second}, even for macroscopic black holes, the derivation of Hawking radiation via Bogoliubov transformations traces field modes from the asymptotic future back to the asymptotic past. This procedure involves modes that, near the black hole's formation event, correspond to arbitrarily high (trans-Planckian) frequencies in the freely falling frame. The \emph{trans-Planckian problem} questions whether the existence and precise thermal spectrum of Hawking radiation depend sensitively on unknown physics at or above the Planck scale \(M_P\). Arguments for robustness include analogue models (e.g., Unruh's hydrodynamic sound analogy) and the Parikh-Wilczek tunneling formalism, which derive a Hawking-like flux without following trans-Planckian frequencies. Conversely, sensitivity is indicated in frameworks with modified high-energy dispersion relations, such as \(\omega^2 = k^2 + \xi\, k^4/M_P^2 + \mathcal{O}(k^{5}) \), where the coefficient \(\xi\) can alter the late-time spectrum or introduce non-thermal corrections. The issue remains open because a complete ultraviolet completion of gravity is required to settle which class of models captures the correct physics near the horizon.

\textbf{Third}, the semiclassical approximation neglects backreaction: the effect of the radiated energy-momentum on the spacetime geometry itself. For the slow evaporation of large black holes, this is an excellent approximation initially, but as the black hole loses significant mass, backreaction effects must become important. We lack a consistent framework for including these effects throughout the entire evaporation process.

\subsection{Quantum Field Theory in Curved Spacetime}

The technical machinery underlying Hawking's calculation is quantum field theory in curved spacetime (QFTCS), which extends standard quantum field theory from Minkowski space to arbitrary curved backgrounds. In flat spacetime, the notion of particles is unambiguous because we have a preferred notion of time translation symmetry and positive frequency modes. In curved spacetime, particularly in spacetimes without timelike Killing vectors, the notion of ``particle'' becomes observer-dependent and basis-dependent.

The key technical tool is the Bogoliubov transformation, which relates different choices of mode decompositions for the quantum field. If we quantize a field using mode functions $\{u_i\}$ (defining creation and annihilation operators $\hat{a}_i^\dagger$, $\hat{a}_i$) and alternatively using mode functions $\{v_j\}$ (defining $\hat{b}_j^\dagger$, $\hat{b}_j$), these operators are related by:
\begin{equation}\label{eq:bogoliubov}
\hat{b}_j = \sum_i \left(\alpha_{ji} \hat{a}_i + \beta_{ji} \hat{a}_i^\dagger\right),
\end{equation}
where the Bogoliubov coefficients $\alpha$ and $\beta$ are determined by the overlap integrals of the mode functions. The crucial point is that a state defined as vacuum with respect to one set of modes ($\hat{a}_i|0\rangle_a = 0$ for all $i$) will generally contain particles from the perspective of the other mode decomposition. The expectation value $\langle 0|_a \hat{b}_j^\dagger \hat{b}_j |0\rangle_a$ gives the particle number in mode $j$ as measured by the second observer.

For black hole spacetimes, Hawking's calculation compares ``in'' modes defined in the distant past (before the black hole formed) with ``out'' modes defined in the asymptotic future (long after formation). The ``in'' vacuum, which represents empty space before the collapse, appears to distant future observers as a thermal bath of particles. This is the origin of Hawking radiation, and its thermal character with temperature $T_H = \hbar\kappa/(2\pi k_B c)$ emerges from the specific form of the Bogoliubov coefficients, which yield a Planck distribution.

However, this derivation rests on several assumptions that merit careful scrutiny. The choice of ``in'' and ``out'' vacuum states, while physically motivated, is not unique. Different observer-dependent vacuum choices would yield different particle content. Furthermore, the calculation assumes eternal black holes or formation scenarios with specific asymptotic behavior. Whether these results apply universally to all black hole formation scenarios remains a subtle question that has generated significant technical literature.

\subsection{The Information Paradox: A Fundamental Tension}

Hawking's 1975 calculation~\cite{Hawking1975} led to a disturbing conclusion: if we trace the quantum state of matter as it collapses to form a black hole and subsequently evaporates via Hawking radiation, a pure quantum state appears to evolve into a mixed thermal state. This violates \emph{unitarity}, the fundamental principle of quantum mechanics that evolution preserves the purity of quantum states. Specifically, if we start with a pure state $|\psi\rangle$ with density matrix $\rho_{\text{initial}} = |\psi\rangle\langle\psi|$ (with $\text{Tr}(\rho^2) = 1$), and the system evolves to a thermal mixed state $\rho_{\text{final}}$ with $\text{Tr}(\rho^2) < 1$, then information has been destroyed~\cite{Hawking1976}.

This information paradox is not merely a technical puzzle but reflects a deep incompatibility between three cherished principles:

\textbf{First}, unitarity of quantum mechanics requires that evolution is described by unitary operators $U$ such that $\rho_{\text{final}} = U \rho_{\text{initial}} U^\dagger$, which preserves $\text{Tr}(\rho^2) = 1$.

\textbf{Second}, the equivalence principle of general relativity requires that local physics at the event horizon should appear ordinary to an infalling observer. Nothing special should happen at the horizon from the perspective of someone crossing it.

\textbf{Third}, locality and causality in quantum field theory require that information cannot propagate faster than light. Information that crosses the horizon should not be accessible to external observers.

Hawking's calculation respects the second and third principles but appears to violate the first. Various proposed resolutions prioritize these principles differently: some preserve unitarity at the cost of modifying physics at the horizon (firewalls, fuzzballs), others maintain horizon smoothness but invoke non-local information transfer, and still others question whether the paradox is properly formulated given our ignorance of quantum gravity.

This tension has not been resolved despite nearly five decades of research, though recent developments using gravitational path integrals have provided new perspectives that we will examine critically in later sections. What is crucial to recognize here is that black hole thermodynamics, despite its remarkable internal consistency within the semiclassical framework, points toward an incompleteness in our current theoretical understanding. The field is not characterized by settled answers but by a spectrum of competing proposals, each abandoning or modifying some aspect of our current theoretical framework.

\section{Bekenstein and the Rise of Black Hole Entropy}
\label{sec:bekenstein_all}

Bekenstein's work in the 1970s established the conceptual foundation for black hole thermodynamics. His key insight was that black holes must possess entropy to preserve the Second Law when matter falls into them. However, his original proposal left the precise numerical coefficient undetermined and relied on heuristic arguments that, while physically insightful, lacked the rigorous foundation that Hawking's subsequent work would provide.

\subsection{Common Entropy Plus Black-Hole Entropy Never Decreases (1972)}
\label{sec:bek1972}

In his 1972 paper~\cite{Bekenstein1972}, Bekenstein introduces the concept of black hole entropy and reformulates the second law of thermodynamics. He proposes that black holes possess entropy proportional to their surface area. The correct, dimensionally consistent form (in SI units) is:
\begin{equation}\label{eq:bek1972_entropy}
S_{\text{bh}} = \eta \, k_B \frac{A}{L_P^2},
\end{equation}
where $A$ is the horizon area, $k_B$ is Boltzmann's constant, $L_P = \sqrt{\hbar G / c^3}$ is the Planck length, and $\eta$ is a dimensionless constant of order unity to be determined. Bekenstein's \emph{Generalized Second Law} (GSL) states that the sum of ordinary entropy outside black holes and black hole entropy never decreases:
\begin{equation}\label{eq:gsl_bek1972}
\delta(S_{\text{ext}} + S_{\text{bh}}) \geq 0.
\end{equation}

Bekenstein addresses two apparent violations of standard thermodynamics: entropy disappearance when matter falls into black holes, and Geroch's \emph{gedankenexperiment} for converting heat entirely into work using a black hole. For black-body radiation with temperature $T \gg \hbar/(k_B M)$ falling into an extreme Kerr black hole, Bekenstein derives the minimum area increase as:
\begin{equation}\label{eq:bek1972_deltaA}
(\Delta A)_{\text{min}} = 8\pi(2-\sqrt{3}) \frac{G}{c^4} M E,
\end{equation}
where $M$ is the black hole mass and $E$ is the energy of the infalling radiation. This calculation assumes a quasi-reversible process where the radiation is lowered slowly into the black hole, an idealization that may not be physically realizable. The corresponding entropy increase is:
\begin{equation}\label{eq:bek1972_deltaS}
(\Delta S_{\text{bh}})_{\text{min}} = \eta \, k_B \frac{(\Delta A)_{\text{min}}}{L_P^2}.
\end{equation}
The decrease in exterior entropy satisfies $|\Delta S_{\text{ext}}| \ll k_B M E / \hbar$, ensuring overall entropy increase when $\eta \sim 1$.

\textbf{Critical assessment}: Bekenstein's 1972 argument is heuristic rather than rigorous. The specific numerical factor in Eq.~\eqref{eq:bek1972_deltaA} depends on the details of the idealized process, and the assumption that such quasi-reversible processes are possible for black holes is non-trivial. Furthermore, the order-of-magnitude estimate $\eta \sim 1$ leaves the precise value undetermined. Bekenstein's great achievement was conceptual: recognizing that black holes must carry entropy to avoid thermodynamic paradoxes. The precise formulation awaited Hawking's quantum mechanical treatment.

\subsection{Entropy-Area Relation and Information Loss (1973)}
\label{sec:bek1973}

Bekenstein refines his proposal in 1973~\cite{Bekenstein1973}, arguing that quantum mechanics is essential to resolve the dimensional mismatch between area (dimension length$^2$) and entropy (dimensionless in natural units). He proposes the entropy-area relation (in SI units):
\begin{equation}\label{eq:bek1973_entropy}
S_{\text{bh}} = \frac{\alpha k_B c^3}{4\hbar G} A,
\end{equation}
where $\alpha$ is a dimensionless constant that Bekenstein estimates to be of order unity but cannot determine precisely from his arguments. This corresponds to $\eta = \alpha/4$ in Eq.~\eqref{eq:bek1972_entropy}. The GSL is formulated as:
\begin{equation}\label{eq:bek1973_GSL}
\delta(S_{\text{bh}} + S_{\text{ext}}) \geq 0.
\end{equation}

Bekenstein shows that capturing a particle with Compton wavelength $\lambda_C = \hbar/(mc)$ leads to a minimal area increase $\Delta A_{\text{min}} \sim L_P^2$, which he associates with losing one bit of information, corresponding to an entropy increase of $\Delta S \sim k_B \ln 2$. This argument suggests $\alpha \sim 1$, but the precise coefficient remains undetermined.

\textbf{Critical assessment}: Bekenstein's 1973 paper makes progress by identifying the role of quantum mechanics in providing the dimensional scales ($\hbar$) necessary to relate area to entropy. However, his information-theoretic arguments, while suggestive, are not quantitatively rigorous. The claim that minimal area increase corresponds to losing exactly one bit involves assumptions about the quantum nature of horizon area that go beyond what can be justified from semiclassical considerations alone. The precise value $\alpha = 1$ (or equivalently $\eta = 1/4$) would require Hawking's calculation to establish definitively.

\section{Hawking: From Skepticism to Revelation}
\label{sec:hawking_all}

Hawking initially rejected Bekenstein's thermodynamic interpretation but later, through quantum field theory in curved spacetime, provided its physical foundation and determined the precise numerical coefficients that Bekenstein's heuristic arguments could not fix.

\subsection{The Four Laws and Initial Skepticism (1973)}
\label{sec:bard1973}

Bardeen, Carter, and Hawking~\cite{Bardeen1973} established four laws structurally analogous to thermodynamics, which we presented in Section 2.1, Eq.~\eqref{eq:first_law_bh}. Hawking initially considered this analogy purely mathematical~\cite{Hawking1973}, arguing that without an emission mechanism, black holes could not be true thermodynamic systems. He explicitly stated that if black holes had temperature, they would need to radiate, but classical general relativity provided no such mechanism. This perspective changed dramatically with his discovery of quantum particle creation~\cite{Hawking1974}.

\subsection{Quantum Fields in Curved Spacetime and Hawking Radiation (1974)}
\label{sec:haw1974}

Hawking's 1974 breakthrough~\cite{Hawking1974} applied quantum field theory to Schwarzschild black holes formed by gravitational collapse. Using the Bogoliubov transformation framework introduced in Section 2.3, Eq.~\eqref{eq:bogoliubov}, he showed that vacuum fluctuations near the event horizon lead to particle creation with a thermal spectrum. The Hawking temperature is:
\begin{equation}\label{eq:hawking_temp}
T_{\text{BH}} = \frac{\hbar \kappa}{2\pi c k_B},
\end{equation}
where $\kappa$ is the surface gravity. For a Schwarzschild black hole, $\kappa = c^4/(4GM)$, yielding:
\begin{equation}\label{eq:hawking_temp_schwarzschild}
T_{\text{BH}} = \frac{\hbar c^3}{8\pi G M k_B}.
\end{equation}
For a solar-mass black hole ($M \approx 2 \times 10^{30}$ kg), this gives $T_{\text{BH}} \approx 6.2 \times 10^{-8}$ K, far below the cosmic microwave background temperature of 2.7 K, rendering such black holes undetectable via their thermal radiation with current technology.

The emission probability follows a distribution:
\begin{equation}\label{eq:hawking_spectrum}
\Gamma(\omega) = \frac{1}{e^{2\pi \omega/\kappa} \mp 1},
\end{equation}
with $-$ for bosons and $+$ for fermions. This is precisely a Planck distribution with temperature $T_{\text{BH}}$ given by Eq.~\eqref{eq:hawking_temp}, confirming the thermal nature of the radiation.

The emission of this radiation leads to mass loss at a rate:
\begin{equation}\label{eq:mass_loss_rate}
\frac{dM}{dt} = -\frac{\hbar c^4}{15360\pi G^2 M^2},
\end{equation}
where the numerical coefficient depends on the number of particle species and their spin. This yields an evaporation timescale:
\begin{equation}\label{eq:evaporation_time}
\tau_{\text{evap}} \sim \frac{G^2 M^3}{\hbar c^4} \approx 2.1 \times 10^{67} \left(\frac{M}{M_\odot}\right)^3 \text{ years}.
\end{equation}

\textbf{Critical assessment}: Hawking's 1974 calculation was revolutionary, but its limitations must be acknowledged. The derivation assumes: (1) the semiclassical approximation remains valid throughout, (2) backreaction effects are negligible, (3) the background geometry is exactly Schwarzschild, (4) the calculation of Bogoliubov coefficients using geometric optics approximation near the horizon is accurate. The first two assumptions certainly break down during late-stage evaporation. The trans-Planckian problem, discussed in Section 2.2, raises questions about whether the result depends on unknown high-energy physics. Despite these limitations, the prediction of thermal radiation with temperature inversely proportional to mass is remarkably robust across different calculation methods.

\subsection{Particle Creation by Black Holes (1975)}
\label{sec:haw1975}

In his comprehensive 1975 paper~\cite{Hawking1975}, Hawking solidifies the thermodynamic picture and determines the precise entropy formula. Comparing the first law of black hole mechanics, Eq.~\eqref{eq:first_law_bh}, with the thermodynamic first law using the temperature from Eq.~\eqref{eq:hawking_temp}, he finds that consistency requires:
\begin{equation}\label{eq:hawking_entropy}
S_{\text{BH}} = \frac{k_B c^3}{4\hbar G} A = \frac{k_B A}{4 L_P^2}.
\end{equation}
This fixes Bekenstein's undetermined coefficient: $\eta = 1/4$ in Eq.~\eqref{eq:bek1972_entropy} and $\alpha = 1$ in Eq.~\eqref{eq:bek1973_entropy}. The Generalized Second Law becomes:
\begin{equation}\label{eq:GSL_final}
\delta\left(S_{\text{ext}} + \frac{k_B c^3}{4\hbar G} A\right) \geq 0.
\end{equation}

For rotating or charged black holes, the emission spectrum includes chemical potentials associated with angular momentum and charge:
\begin{equation}\label{eq:rotating_spectrum}
\Gamma = \frac{1}{\exp\left[2\pi(\omega - m\Omega - eQ\Phi)/\kappa\right] \mp 1},
\end{equation}
where $m$ is the azimuthal quantum number, $\Omega$ is the angular velocity of the horizon, $e$ is the particle charge, $Q$ is the black hole charge, and $\Phi$ is the electrostatic potential at the horizon. This is a grand canonical distribution, indicating that black holes emit particles preferentially in modes that carry away angular momentum and charge, eventually approaching a Schwarzschild configuration.

\textbf{Critical assessment}: The 1975 paper establishes black hole thermodynamics as a consistent semiclassical framework, but it simultaneously reveals the information paradox. Hawking explicitly argues that information is lost during black hole evaporation, violating quantum unitarity~\cite{Hawking1976}. This conclusion, which Hawking maintained for decades~\cite{Hawking2005}, creates a fundamental tension with quantum mechanics that remains controversial. The consistency of the thermodynamic laws within the semiclassical approximation does not resolve the deeper quantum mechanical issues.

\section{Synthesis and the Generalized Second Law}
\label{sec:synthesis}

The combined work of Bekenstein and Hawking established black hole thermodynamics on firm ground within the semiclassical approximation. The GSL, as given in Eq.~\eqref{eq:GSL_final}, can be written compactly as:
\begin{equation}\label{eq:gsl_compact}
\delta S_{\text{total}} = \delta\left(S_{\text{ext}} + \frac{k_B A}{4L_P^2}\right) \geq 0,
\end{equation}
where $L_P^2 = \hbar G/c^3$. This law resolves the apparent thermodynamic paradoxes that motivated Bekenstein's original work: when matter falls into a black hole, the decrease in external entropy is compensated by an increase in black hole entropy, preserving the total.

However, several caveats must be emphasized. First, the GSL has been proven to hold for semiclassical processes where quantum field theory in curved spacetime is valid, but we cannot be certain it holds when quantum gravitational corrections become important. Second, the GSL does not resolve the information paradox --- indeed, the information paradox can be formulated as a tension between the GSL (which suggests irreversible processes) and quantum unitarity (which requires reversibility). Third, the physical interpretation of black hole entropy remains unclear: what are the microstates being counted by $S_{\text{BH}} = k_B A/(4L_P^2)$?

\subsection{Statistical Mechanical Interpretation (1975)}
\label{sec:bek1975_stat}

Bekenstein's 1975 work~\cite{Bekenstein1975} sought a statistical foundation for black hole entropy. He proposed that $S_{\text{bh}}$ counts microstates of the horizon, with the coefficient $\eta = 1/4$ (now determined by Hawking's calculation) entering the counting. The \emph{Generalized Maximum Entropy Principle} (GMEP) extends standard maximum entropy methods to include black hole degrees of freedom, yielding probability distributions for black hole parameters. This suggests that macroscopic black hole parameters (mass, angular momentum, charge) may be distributed according to statistical ensembles rather than having sharp values.

\textbf{Critical assessment}: While conceptually appealing, Bekenstein's statistical mechanical interpretation raises more questions than it answers. What physical degrees of freedom constitute these microstates? Are they localized at the horizon or distributed throughout the black hole interior? How do we reconcile a statistical mechanical description with the no-hair theorem, which suggests black holes have no internal structure? These questions would motivate decades of research in string theory and loop quantum gravity, which we examine in the next section.

\section{Impact and Open Problems}
\label{sec:impact}

The Bekenstein-Hawking entropy has profoundly influenced modern theoretical physics, catalyzing developments in quantum gravity, holography, and quantum information. However, many of these developments remain speculative, and it is crucial to distinguish between established results and ongoing research programs.

\subsection{Holographic Principle and AdS/CFT}

The entropy-area relation inspired the holographic principle~\cite{tHooft1993,Susskind1995}, which posits that the information content of a volume is encoded on its boundary. The maximum entropy in a spatial region is proportional to the boundary area rather than the volume, with the Bekenstein-Hawking formula providing the proportionality constant. This principle finds a concrete realization in the AdS/CFT correspondence~\cite{Maldacena1997}, where a gravitational theory in $(d+1)$-dimensional anti-de Sitter (AdS) space is equivalent to a conformal field theory (CFT) without gravity on its $d$-dimensional boundary. This duality provides a non-perturbative definition of quantum gravity in AdS spacetimes and has been used to compute black hole entropy microscopically in certain cases.

Within AdS/CFT, the Ryu-Takayanagi formula~\cite{Ryu2006} relates the entanglement entropy of a subregion in the CFT to the area of a minimal surface in the bulk AdS geometry:
\begin{equation}\label{eq:ryu_takayanagi}
S_{\text{EE}} = \frac{A_{\text{min}}}{4 G \hbar / c^3} + S_{\text{bulk}},
\end{equation}
where $A_{\text{min}}$ is the area of the extremal surface homologous to the boundary subregion, and $S_{\text{bulk}}$ accounts for bulk entanglement entropy. For static black holes in AdS, this reproduces the Bekenstein-Hawking entropy when the subregion encompasses the entire boundary.

\textbf{Critical assessment}: While AdS/CFT provides a concrete framework for studying quantum gravity, its applicability to realistic black holes in asymptotically flat or de Sitter spacetimes remains unproven. The correspondence is rigorously established only for specific supersymmetric theories and has not been derived from first principles. Moreover, computations often rely on the large-$N$ limit (where $N$ is the rank of the gauge group in the CFT), which may not capture finite-$N$ effects relevant to real-world gravity. The holographic principle, while inspired by black hole entropy, is a conjecture whose general validity beyond AdS/CFT is contested.

\subsection{Microstate Counting in String Theory and Loop Quantum Gravity}

Attempts to derive black hole entropy from microstate counting have been pursued in string theory and loop quantum gravity, providing candidate statistical mechanical foundations for $S_{\text{BH}}$.

In string theory, Strominger and Vafa~\cite{Strominger1996} computed the entropy of extremal five-dimensional black holes by counting BPS-saturated D-brane configurations at weak coupling, reproducing exactly:
\begin{equation}\label{eq:strominger_vafa}
S_{\text{BH}} = \frac{k_B A}{4 L_P^2},
\end{equation}
where the area $A$ is evaluated in the strong-coupling regime where the D-brane system becomes a black hole. This matching relies on supersymmetry to protect the counting from quantum corrections as the coupling is varied. Subsequent work extended this approach to certain near-extremal~\cite{Maldacena1997b} and non-extremal cases, though a general microscopic description for astrophysical Kerr black holes remains elusive.

In loop quantum gravity (LQG), Rovelli~\cite{Rovelli1996} and Ashtekar et al.~\cite{Ashtekar1998} model the horizon as a surface punctured by spin network edges, with entropy arising from the degeneracy of boundary spin configurations satisfying the isolated horizon boundary conditions. The entropy is:
\begin{equation}\label{eq:lqg_entropy}
S_{\text{BH}} = \frac{k_B \gamma A}{4 L_P^2},
\end{equation}
where $\gamma$ is the Immirzi parameter, a free constant fixed to $\gamma \approx 0.237$ to match the Bekenstein-Hawking formula.

\textbf{Critical assessment}: Both approaches successfully reproduce the area law up to numerical factors, but neither provides a complete microscopic description for general non-extremal, evaporating black holes. String theory microstate counting is largely limited to supersymmetric or near-extremal cases, where evaporation is suppressed, and extrapolation to realistic astrophysical black holes requires unverified assumptions. The \emph{fuzzball} proposal~\cite{Mathur2005} attempts to extend these ideas to horizon-scale structure, but its dynamical consequences for formation and evaporation are not fully developed. LQG's counting depends on the choice of boundary conditions and the Immirzi parameter, whose physical origin remains unclear. Neither framework has yet produced a first-principles derivation of Hawking radiation or a complete dynamical description of evaporation that resolves the information paradox.

\subsection{Recent Developments: Gravitational Path Integrals and Replica Wormholes}

Recent progress on the information paradox has employed Euclidean gravitational path integrals to compute the entropy of Hawking radiation. Penington~\cite{Penington2020} and Almheiri et al.~\cite{Almheiri2019} introduced the ``island'' prescription, where the entanglement entropy of the radiation is given by:
\begin{equation}\label{eq:island_formula}
S_{\text{rad}} = \min_{\text{islands}} \left[ \frac{k_B A(\partial I)}{4 L_P^2} + S_{\text{matter}}(R \cup I) \right],
\end{equation}
where $R$ is the radiation region outside the black hole, $I$ is a bulk ``island'' region, and the minimum is taken over possible island configurations. This formula, motivated by the quantum extremal surface prescription in AdS/CFT, reproduces the Page curve~\cite{Page1993}, where the radiation entropy initially increases but then decreases after the Page time $t_{\text{Page}} \sim M^3$ (in Planck units), a behavior consistent with unitary evolution.

The technical justification involves replica wormholes in the gravitational path integral~\cite{Almheiri2020}, which are saddle points in the Euclidean integral for the R\'enyi entropy $\text{Tr}(\rho^n)/[\text{Tr}(\rho)]^n$ that connect multiple replicas of the spacetime. When these contributions dominate, they lead to an entanglement entropy for the radiation that follows the Page curve. However, reproducing the Page curve for the entanglement entropy is a necessary but not sufficient condition for proving full unitary evolution of the quantum state. The prescription relies on a non-local geometric structure (the island) whose precise microscopic origin and compatibility with local effective field theory remain subjects of active debate.

This approach exists within a broader landscape of proposed resolutions to the information paradox. The \emph{fuzzball} proposal in string theory posits that black hole microstates are horizonless, quantum-gravitational objects, preventing information loss from the outset~\cite{Mathur2005}. The \emph{firewall} argument~\cite{Almheiri2013} highlighted the tension between unitarity and smooth horizons, suggesting that dramatic departures from semiclassical physics at the horizon might be inevitable. Subsequent work has sought to reconcile these perspectives, with some arguing that islands effectively encode the interior degrees of freedom in the radiation, evading the need for firewalls~\cite{Penington2020}. Nevertheless, significant technical critiques have been raised regarding the consistency and interpretation of the island prescription, particularly concerning its reliance on Euclidean path integrals over disconnected topologies and the challenge of deriving it from a unitary, Lorentzian formulation~\cite{Bousso2020}.

\textbf{Critical assessment}: While mathematically elegant, the island/replica wormhole approach relies on several unproven assumptions. The inclusion of wormhole geometries in the gravitational path integral is a formal step whose physical justification in Lorentzian signature remains unclear. The consensus is limited: while the reproduction of the Page curve is widely acknowledged, whether this constitutes a resolution of the information paradox is debated. Critics note that the mechanism for information recovery remains obscure, as the island prescription does not specify how individual quantum states are transferred from the black hole interior to the radiation. Furthermore, the trans-Planckian problem and backreaction effects are not systematically incorporated, and the approach has been primarily developed in Jackiw–Teitelboim (JT) gravity and AdS settings, with its extension to asymptotically flat spacetimes still tentative.

\section{Conclusion: Established Knowledge and Persistent Challenges}

Black hole thermodynamics stands as one of the most profound achievements at the interface of general relativity and quantum mechanics. The core results --- the Hawking temperature $T_{\text{BH}} = \hbar \kappa / (2\pi c k_B)$, the Bekenstein-Hawking entropy $S_{\text{BH}} = k_B A / (4 L_P^2)$, and the Generalized Second Law --- are firmly established within the semiclassical approximation for macroscopic black holes. These follow rigorously from quantum field theory in curved spacetime and are robust against perturbations in the calculation methods.

However, substantial limitations persist. The semiclassical framework breaks down for Planck-scale black holes, where quantum gravity effects dominate, and we lack experimental verification due to the minuscule radiation from astrophysical black holes. The information paradox remains unresolved: while recent proposals like islands and replica wormholes suggest unitarity can be preserved, they rely on speculative extensions of semiclassical gravity and do not command universal consensus. The trans-Planckian problem questions the sensitivity to unknown ultraviolet physics, and backreaction effects during evaporation are not fully incorporated.

In summary, the thermodynamic properties of black holes are well-understood semiclassically but speculative beyond that regime. Future progress requires a complete theory of quantum gravity, potentially testable through multimessenger astronomy or analog systems, but for now, the field is characterized by deep insights coexisting with fundamental incompleteness.

\onecolumn
\appendix
\section*{Appendix A: Notation Conventions and Equation Translation}
\label{app:notation}

\subsection*{Motivation and Scope}

This article traces the historical development of black hole thermodynamics from 1972 to 1975, presenting key equations in the original notation used by Bekenstein and Hawking. This choice preserves the authenticity of their respective derivations and reflects the conceptual evolution of the field. However, the different conventions for physical constants, units, and notation can hinder direct comparison. This appendix provides a comprehensive ``translation dictionary'' between the major conventions used in the original papers and the modern, dimensionally consistent SI notation employed in the main text's analysis.

\subsection*{Conventions Dictionary}

The table below summarizes the primary conventions. Note that both Bekenstein and Hawking often used ``geometric units'' where $G = c = \hbar = k_B = 1$, but their explicit equations sometimes restored these constants for clarity. The main text uses SI units throughout for dimensional clarity and consistency.

\begin{table}[H]
\centering
\caption{Notation conventions in original papers and their SI equivalents.}
\label{tab:conventions}
\begin{tabular}{p{0.22\linewidth} p{0.22\linewidth} p{0.22\linewidth} p{0.25\linewidth}}
\toprule
\textbf{Physical Concept} & \textbf{Bekenstein (1972--73)} & \textbf{Hawking (1974--75)} & \textbf{Modern SI (Main Text)} \\
\midrule
Speed of Light & Often implicit, sometimes $c=1$ & Often $c=1$ & $c$ \\
\hline
Gravitational Constant & $G$ & $G$ (sometimes $\kappa$) & $G$ \\
\hline
Planck's Constant & $\hbar$ or $h$ & $\hbar$ (often $h=2\pi\hbar$) & $\hbar$ \\
\hline
Boltzmann's Constant & $k$ & $k$ & $k_B$ \\
\hline
Planck Length & Not explicitly defined, but $L_P^2 \propto \hbar G/c^3$ & Not always explicit & $L_P = \sqrt{\hbar G / c^3}$ \\
\hline
Black Hole Entropy & $S_{\text{bh}} = \eta \, k A / L_P^2$ (1972) & $S_{\text{BH}} = A/(4 L_P^2)$ (in units with $k_B=1$) & $S_{\text{BH}} = \dfrac{k_B c^3}{4\hbar G} A$ \\
& $S_{\text{bh}} = \dfrac{\alpha k c^3}{4\hbar G} A$ (1973) & & \\
\hline
Horizon Area & $A$ & $A$ & $A$ \\
\hline
Surface Gravity & $\kappa$ (sometimes implicit) & $\kappa$ & $\kappa$ \\
\hline
Hawking Temperature & Not derived (pre-Hawking) & $T_{\text{BH}} = \kappa/(2\pi)$ (with $k_B=1, \hbar=1$) & $T_{\text{BH}} = \dfrac{\hbar \kappa}{2\pi c k_B}$ \\
\hline
Mass Loss Rate & Not derived & $\dfrac{dM}{dt} \propto -1/M^2$ (in geometric units) & $\dfrac{dM}{dt} = -\dfrac{\hbar c^4}{15360\pi G^2 M^2}$ \\
\bottomrule
\end{tabular}
\end{table}

\subsection*{Key Equation Translations}

Below, we explicitly translate the most important equations from their original forms (as they appear in the cited papers) to the unified SI notation used in the main text for comparison.

\paragraph{1. Black Hole Entropy}
\begin{itemize}
  \item \textbf{Bekenstein (1972)} originally wrote~\cite{Bekenstein1972}: $S_{\text{bh}} = \eta \, k \, A / L_P^2$, with $\eta \sim 1$ and $L_P^2$ understood as $\hbar G/c^3$. In SI form (main text Eq.~\eqref{eq:bek1972_entropy}):
        \[ S_{\text{bh}} = \eta \, k_B \frac{A}{L_P^2}, \quad L_P^2 = \frac{\hbar G}{c^3}. \]
    \item \textbf{Bekenstein (1973)} refined this to~\cite{Bekenstein1973}: $S_{\text{bh}} = \dfrac{\alpha k c^3}{4\hbar G} A$ (main text Eq.~\eqref{eq:bek1973_entropy}). Setting $\alpha=1$ and using $k_B$ gives the SI form.
    \item \textbf{Hawking (1975)} established the exact coefficient~\cite{Hawking1975}: $S_{\text{BH}} = A/(4 L_P^2)$ in Planck units ($\hbar = G = c = k_B = 1$). Restoring all constants yields (main text Eq.~\eqref{eq:hawking_entropy}):
        \[ S_{\text{BH}} = \frac{k_B c^3}{4\hbar G} A = \frac{k_B A}{4 L_P^2}. \]
        This fixes $\eta = 1/4$ and $\alpha = 1$ in Bekenstein's earlier expressions.
\end{itemize}

\paragraph{2. Hawking Temperature}
\begin{itemize}
    \item \textbf{Hawking (1974)} found the temperature~\cite{Hawking1974} as $T_{\text{BH}} = \kappa/(2\pi)$ in units with $\hbar = c = k_B = 1$. The dimensionally correct SI form (main text Eq.~\eqref{eq:hawking_temp}) is:
        \[ T_{\text{BH}} = \frac{\hbar \kappa}{2\pi c k_B}. \]
        For a Schwarzschild black hole ($\kappa = c^4/(4GM)$), this becomes (main text Eq.~\eqref{eq:hawking_temp_schwarzschild}):
        \[ T_{\text{BH}} = \frac{\hbar c^3}{8\pi G M k_B}. \]
\end{itemize}

\paragraph{3. Generalized Second Law (GSL)}
\begin{itemize}
    \item \textbf{Bekenstein (1972--73)} formulated the GSL as $\delta(S_{\text{ext}} + S_{\text{bh}}) \geq 0$, with $S_{\text{bh}}$ as above~\cite{Bekenstein1972,Bekenstein1973}.
    \item \textbf{Hawking (1975)} confirmed this with the precise entropy formula~\cite{Hawking1975}. The unified SI statement (main text Eq.~\eqref{eq:gsl_compact}) is:
        \[
        \delta S_{\text{total}} = \delta\left( S_{\text{ext}} + \frac{k_B A}{4 L_P^2} \right) \geq 0.
        \]
\end{itemize}

\paragraph{4. Mass Loss Rate}
\begin{itemize}
    \item \textbf{Hawking (1974)} gave the power emitted in geometric units~\cite{Hawking1974}. The full SI expression (main text Eq.~\eqref{eq:mass_loss_rate}) is:
        \[ \frac{dM}{dt} = -\frac{\hbar c^4}{15360\pi G^2 M^2}, \]
        where the numerical factor assumes only massless photons and gravitons; additional particle species increase the rate.
\end{itemize}

\subsection*{Note on Historical Presentation}

The main text presents equations in a hybrid manner: when first introducing a historical result, we quote it in a form that closely mirrors the original author's notation while ensuring dimensional consistency (typically by restoring fundamental constants implicitly set to unity). Immediately following, we often provide the modern SI equivalent or discuss the translation. This approach allows the reader to appreciate the original context while maintaining the ability to compare results across different papers.

We emphasize that the apparent inconsistencies between equations are not errors but reflections of the evolving and sometimes divergent notational choices in the pioneering literature. The translation provided here and the careful presentation in the main text aim to bridge these differences without sacrificing historical fidelity.

\twocolumn

\end{document}